\documentclass[conference]{IEEEtran}
\IEEEoverridecommandlockouts
\usepackage{cite}
\usepackage{amsmath,amssymb,amsfonts}
\usepackage{algorithmic}
\usepackage{graphicx}
\usepackage{textcomp}
\usepackage{xcolor}
\usepackage{orcidlink}
\hypersetup{pdfborder={0 0 0}}
\def\BibTeX{{\rm B\kern-.05em{\sc i\kern-.025em b}\kern-.08em
    T\kern-.1667em\lower.7ex\hbox{E}\kern-.125emX}}
\usepackage{cleveref}
\usepackage{float}
\floatstyle{ruled}
\newfloat{listing}{thp}{lop}
\floatname{listing}{Listing}
\begin{document}

\title{Lab-Scale Gantry Crane Digital Twin Exemplar\\
}

\author{\IEEEauthorblockN{Joost Mertens\orcidlink{0000-0002-8148-5024}}
\IEEEauthorblockA{\textit{Cosys-Lab \& Flanders Make} \\
\textit{University of Antwerp}\\
Antwerpen, Belgium \\
joost.mertens@uantwerpen.be
}
\and
\IEEEauthorblockN{Joachim Denil\orcidlink{0000-0002-4926-6737}}
\IEEEauthorblockA{\textit{Cosys-Lab \& Flanders Make} \\
\textit{University of Antwerp}\\
Antwerpen, Belgium \\
joachim.denil@uantwerpen.be
}
}

\maketitle

\begin{abstract}
The research topic of digital twins has attracted a large amount of interest over the past decade. However, publicly available exemplars remain scarce. In the interest of open and reproducible science, in this exemplar paper we present a lab-scale gantry crane and its digital twin. The exemplar comprises both the physical and digital side of the twin system. The physical side consists of the physical crane and its controller. The digital side covers the CAD models and kinematic model of the crane, and provides services for optimal control, historical data logging, data visualization and continuous validation. We used this setup as use case in several previous publications where its functionality was validated. It is publicly available and only relies on other freely available and commonly used software, this way we hope it can be used for future research or education on the topic of digital twins.

\end{abstract}


\section{Introduction}
\label{sec:introduction}

The digital twin paradigm connects an actual system with a digital counterpart consisting of models of parts of the actual systems and services derived therefrom. During operation data flows from the actual to the digital system to feed the twin's services. In turn, the twin's services improve the actual system's workings. While the paradigm is applicable for both physical and digital actual objects~\cite{Paredis-2024}, in practice we see it mainly applied to physical ones~\cite{Cimino-2019,Madubuike-2022}.

Because of this bi-directional interaction, demonstrating research results requires relevant use cases. Such research results could be new synchronization methods, better or faster data streaming, improved services and the like. The exemplar in this paper for example, was used as a demonstrator for the following digital twin related research: continuous validation methods~\cite{Mertens-2023,Mertens-2024}, fault detection methods~\cite{Mertens-2024a} and DarTwin, a notation for system evolution~\cite{Mertens-2024b}.

In our experience, the use cases themselves typically do not contain new research, yet they demand a substantial development effort on the researcher's part. Finding of-the-shelf exemplars in literature is rare, as the use cases in papers are often not described in a reproducible manner, are costly to set up, or are not openly available because they contain intellectual property. Having readily available off-the-shelf exemplars in various domains would help reduce development overhead, and additionally also also help support the generalizability of research results. 

Given the industry relevance of digital twins, education is another field for which well described exemplars can be beneficial, aiding teachers with explaining and demonstrating the digital twin concept, and training engineers with the necessary experience to set up such systems in the real world.

In this paper, we present an openly available digital twin exemplar of a lab-scale gantry crane. It is inspired by cranes in the harbor of Antwerp, which move containers between moored ships and the quay. It is a typical Cyber-Physical System and consists of:

\begin{itemize}
    \item The physical gantry crane including its controller.
    \item The crane's digital twin containing CAD models, services for data logging, data visualization, optimal trajectory generation, and continuous validation.
\end{itemize}

\begin{figure}
    \centering
    \includegraphics[width=0.95\linewidth]{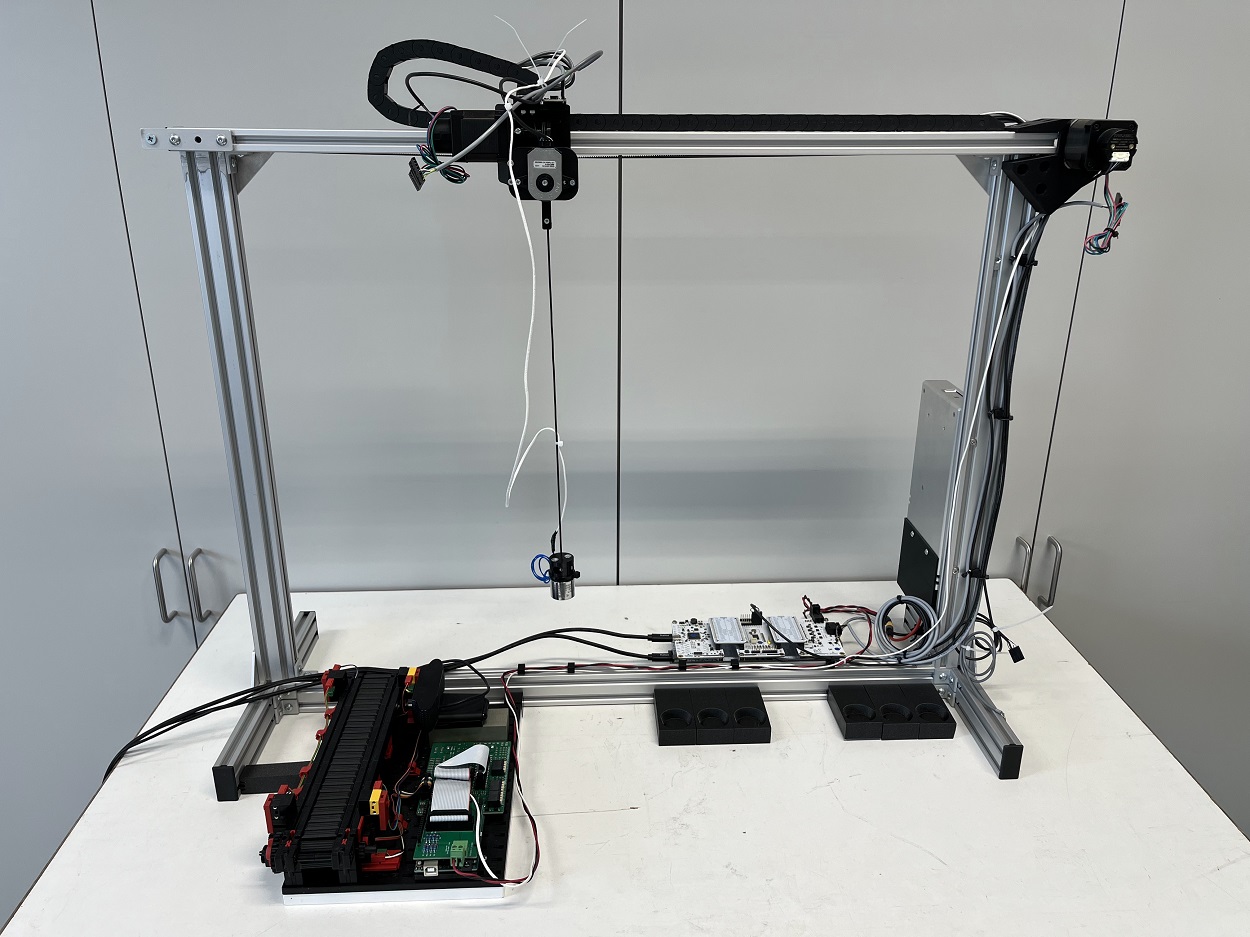}
    \caption{The physical gantry crane.}
    \label{fig:physical-gantry-crane}
\end{figure}

The exemplar is hosted on Cosys-Lab's organizational GitHub\footnote{\url{https://github.com/Cosys-Lab/lab-scale-gantry-crane}}. A picture of the physical crane is shown in \cref{fig:physical-gantry-crane}.

The remainder of this paper is structured as follows: \cref{sec:background} first discusses related exemplars from literature. \Cref{sec:gantry-crane} discusses our exemplar in detail, both architecturally and implementation wise. Afterwards \cref{sec:results} discusses the results and the publications that feature this exemplar. Next, \cref{sec:discussion} discusses some limitations and future extensions. Finally \cref{sec:conclusion} concludes the paper.

\section{Background \& Related Work}
\label{sec:background}

A search with the query ``digital twin exemplar'' on Semantic Scholar, Google Scholar and GitHub's search function yielded other openly available digital twin exemplars with accompanying publications. In what follows, we briefly discuss the most noteworthy ones.

\subsection{ARCHES PiCar-X}

The PiCar-X is a little Raspberry-Pi based robotic car that is used by Barbie and Hasselbring as a demonstrator of various digital twin concepts \cite{Barbie-2024a,Barbie-2024b}. These digital twin concepts\cite{Barbie-2024} were originally used in the ARCHES project on autonomous robotic networks\cite{Barbie-2022}, but the authors wanted to make these results reproducible in a lab-scale setting with an affordable and straightforward example. The exemplar is hosted on GitHub\footnote{\url{https://github.com/cau-se/ARCHES-PiCar-X}} and contains documentation on how to get the exemplar up and running. Emulators are also provided such that the physical system is not necessarily needed. The exemplar features technologies such as Docker containerization to encapsulate the examples and is built on the open ARCHES Digital Twin Framework which is built on top of the ROS (Robot Operation System) middleware.

\subsection{GreenhouseDT}

GreenhouseDT is an exemplar made to facilitate research on self-adaptive digital twins based on a low-cost greenhouse \cite{Kamburjan-2024}. It features behavioral adaptation of both the actual system (the greenhouse) and the digital twin. The exemplar and its accompanying documentation is hosted on GitHub\footnote{\url{https://github.com/smolang/GreenhouseDT}}. It relies on commonly used technologies such as an InfluxDB database, an ActiveMQ message broker and virtual machines for easy installation. The virtual machine images come with prerecorded data.

\subsection{Air Quality Management}

The Air Quality Management exemplar was created to research challenges in the field of Model-Driven digital twin engineering \cite{Govindasamy-2021}. It focuses on a construction engineering case of indoor air quality management in the context of virus spread prevention. Three example applications of the digital twin are provided: runtime visualization, simulation and prediction. The exemplar and instructions on how to set it up are hosted on GitHub\footnote{\url{https://github.com/derlehner/IndoorAirQuality_DigitalTwin_Exemplar}}. It features a Raspberry-Pi with a set of sensors for the actual system and an AutomationML based digital twin \cite{Lehner-2021}. Microsoft Azure tools are used to implement the presented digital twin pipeline.

\subsection{Discussion}

The three exemplars demonstrate the use of digital twins in the robotics, agricultural and construction domains. They commonly use single board computers such as the Raspberry-Pi and containerization or virtualization technologies for easy dissemination of the code. Interestingly, only one of them uses a commercial digital twin framework, while the others are more ad-hoc.

In this exemplar, similar design choices (e.g. use of containerization) are also made, and we utilize a more or less ad-hoc approach rather than using a digital twin  framework. The case itself does stem from a different domain, it is more logistics and cyber-physical systems related, which in our opinion makes it a good addition to these existing exemplars.

\section{Lab Scale Gantry Crane}
\label{sec:gantry-crane}

The lab scale gantry crane is a small gantry crane system inspired by the harbor cranes in the port of Antwerp. It moves small pucks which represents containers from one location to another.

In what follows, we use the C4 Model\footnote{\url{https://c4model.com/}} to visualize and describe the architecture of the system with a context and container diagram. Although it's tailored for software architectures, we still find it applicable for the gantry crane system.

\subsection{Architecture}

\begin{figure}
    \centering
    \includegraphics[width=\linewidth]{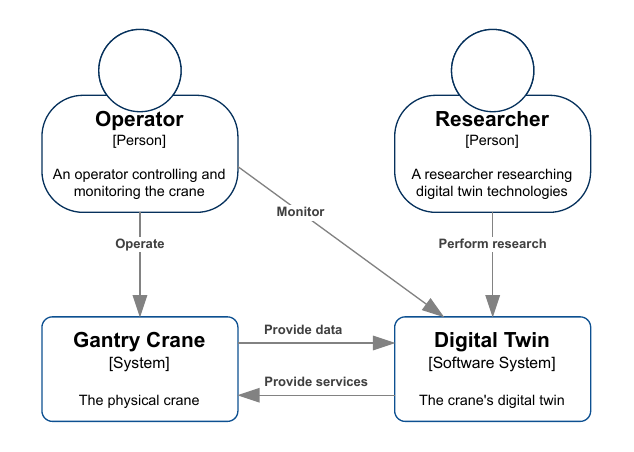}
    \caption{Context diagram of the system.}
    \label{fig:context-diagram}
\end{figure}

\Cref{fig:context-diagram} shows the context diagram of the system. In essence, it shows the high-level abstraction of what a digital twin is, and two actors that interact with the system. The operator is the actor that would operate and monitor this system day to day, whilst the researcher can perform digital twin research on it.

\begin{figure*}
    \centering
    \includegraphics[width=\linewidth]{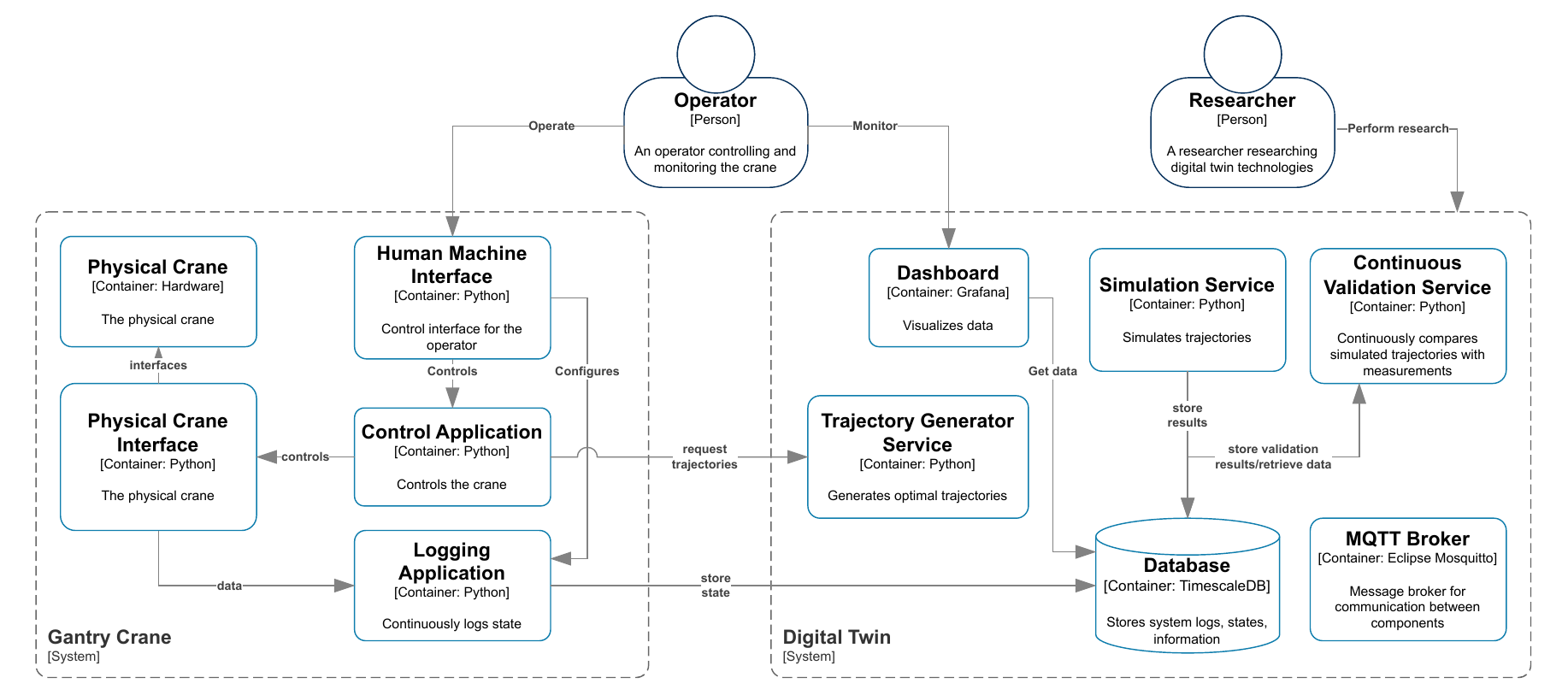}
    \caption{Container diagram of the system.}
    \label{fig:container-diagram}
\end{figure*}

\Cref{fig:container-diagram} shows the container\footnote{Not a Docker container, a container is an abstraction in the C4 model for something that must be there for the system to work.} diagram of the system. We have limited ourselves to the most necessary relationships to keep the diagram readable.

The gantry crane system, which can operate without digital twin if needed, consists of five containers.

\textbf{Physical Crane:} the actual hardware of the crane.

\textbf{Physical Crane Interface:} the interface to the physical crane, a Python program connecting to the controllers of the crane.

\textbf{Control Application:} a Python application that uses the crane interface to control the crane.

\textbf{Logging Application:} a Python application that continuously logs the crane's state to the database.

\textbf{Human Machine Interface:} a Python front-end for the operator to configure and control the crane.

The digital twin consists of six containers. We'll describe them bottom up.

\textbf{Database:} a TimescaleDB database that contains all the data created by the system and its various services.

\textbf{Dashboard:} a Grafana dashboard that visualizes data from the database.

\textbf{Trajectory Generator Service:} a service that utilizes a kinematics model of the crane to generate optimal movement trajectories for the crane. Such optimal trajectories have no residual swinging motion at the end.

\textbf{Simulation Service:} a service that implements a simulation of the crane and it's controllers. It can simulate for example trajectories generated by the Trajectory Generator

\textbf{Continuous Validation Service:} a service that continuously compares measurements of executed trajectories with simulations of those same trajectories. It calculates distance metrics between them and compares the value with an acceptable threshold. In case the threshold is breached operators are notified via the dashboard and can take appropriate action.

\textbf{MQTT Broker:} an MQTT broker for signaling between services.

Finally, we will not discuss the components of each of the containers, but limit ourselves to the \textbf{Physical Crane} to give an overview of the components powering it. We should note that for this physical system, we take a liberal interpretation of what goes in a C4 Model component diagram.

\begin{figure}
    \centering
    \includegraphics[width=\linewidth]{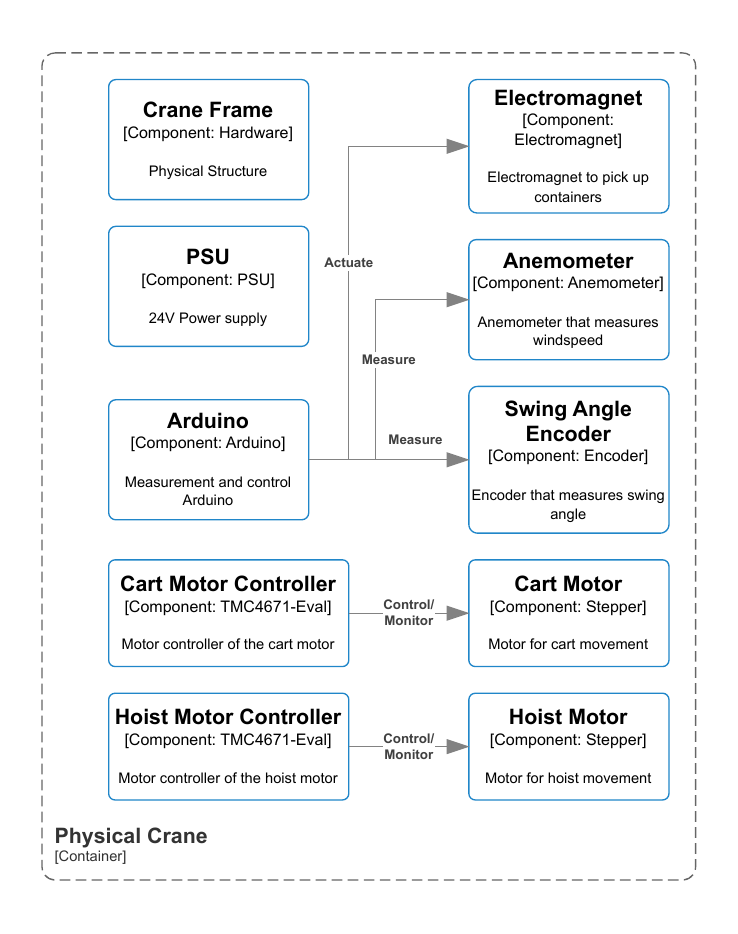}
    \caption{Component Diagram of the Physical Crane.}
    \label{fig:component-diagram}
\end{figure}

\Cref{fig:component-diagram} shows this component diagram, again for brevity and readability only the most necessary relationships are shown. It contains the \textbf{Crane Frame} consisting of all the hardware constituting the physical structure of the crane. Two \textbf{Motors} and associated \textbf{Controllers}, one to control the cart's movement, one to control the hoisting movement. An \textbf{Electromagnet} at the end of the hoist to pick up containers, a \textbf{Swing Angle Encoder} measures the swinging motion of the hoist, and an \textbf{Anemometer} measures the wind speed. An \textbf{Arduino} measures and controls those last three components. Lastly a \textbf{PSU} powers the electronics of the crane.

\subsection{Implementation}
\begin{figure}
    \centering
    \includegraphics[width=0.8\linewidth]{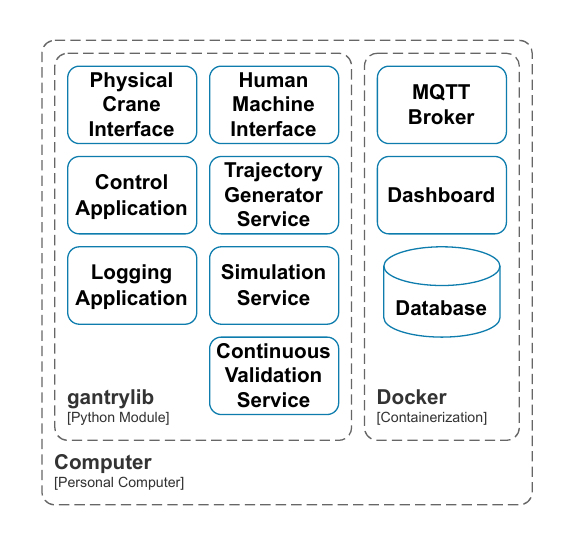}
    \caption{Deployment of components on a personal computer.}
    \label{fig:deployment}
\end{figure}

Regarding the implementation, we wanted to keep the operation of the system as simple as possible, that is, plug in the physical crane to a personal computer, and run all of the software on that computer. \Cref{fig:deployment} shows the distribution of the C4 Containers visually.

For the implementation, we bundled all the Python C4 Containers in a Python module named \textit{gantrylib}. To use this module, it suffices to install it with \textit{pip}. This way it remains easy for researchers to implement additional digital twin services. 

For ease of deployment, the \textbf{Database}, \textbf{MQTT Broker} and \textbf{Dashboard} C4 Containers are containerized with Docker. A docker compose file is provided to bring the containers up and down with a single command. For the first time setup, all of the database and dashboard initialization are also automatically handled by this docker compose file.

Furthermore some of the components warrant a small explanation for their implementation choices. For the database, the choice for TimescaleDB was driven by three factors: (i) it's based on PostgreSQL, allowing for commonly known SQL queries for access. (ii) it's optimized for time series data, which is what is logged. (iii) it's open source. For the dashboard, the choice for Grafana was made because of (i) it's wide support for data sources and (ii) again the fact that it's open source. Lastly, for the MQTT Broker, the ubiquitous and open-source Eclipse Mosquitto was used.

\subsection{Repository Organization}

\Cref{listing:repo} shows the repository organization. The \textbf{BOM} folder contains the Bill-of-Materials. The \textbf{C4-diagrams} folder contains the diagrams used in this paper. The \textbf{CAD} folder contains CAD models of the crane made in Autodesk Fusion360. The \textbf{arcadia} folder contains Arcadia models made in Capella for a lecture on Model-Based Systems Engineering that starred the crane. The \textbf{arduino} folder contains the code to deploy on the measurement and control Arduino, as well as hardware schematics and a board for a custom shield made in KiCad that simplifies connectivity of the sensor to the Arduino. The \textbf{docker} folder contains the docker compose file and all associated first-time setup files for the dashboard, database and MQTT broker. The \textbf{docs} folder contains the documentation made in MkDocs and hosted on GitHub Pages. The \textbf{examples} folder contains basic examples of the gantrylib module. The \textbf{gantrylib} folder then contains the module itself. The \textbf{test} folder contains test associated with the gantrylib module. The \textbf{LICENSE} file contains the MIT license under which the exemplar is published. The \textbf{README} gives a brief overview of the repository and redirects to github pages for the installation. Lastly, \textbf{mkdocs.yml} contains some MkDocs configuration and \textbf{setup.py} contains the metadata of the gantrylib module.

\begin{listing}
\begin{verbatim}
./
|-- BOM/
|-- C4-diagrams
|-- CAD/
|-- arcadia/
|-- arduino/
|-- docker/
|-- docs/
|-- examples/
|-- gantrylib/
|-- tests/
|-- LICENSE
|-- README.md
|-- mkdocs.yml
|-- setup.py
\end{verbatim}   
\caption{Repository organization.}
\label{listing:repo}
\end{listing}

\section{Results}
\label{sec:results}

\begin{figure}
    \centering
    \includegraphics[width=0.95\linewidth]{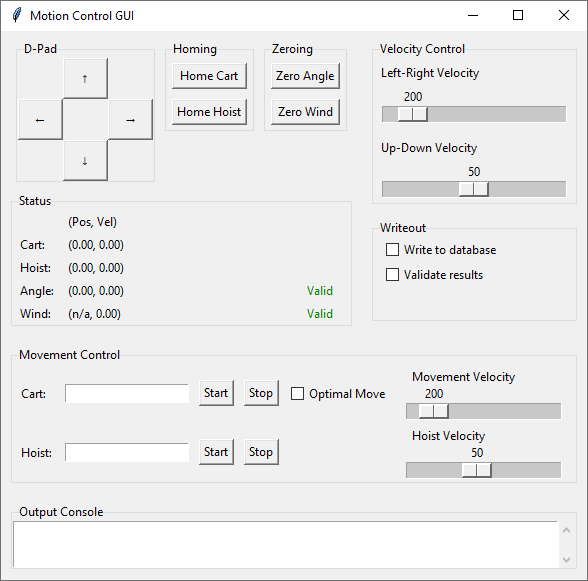}
    \caption{The Human Machine Interface.}
    \label{fig:hmi}
\end{figure}

\begin{figure}
    \centering
    \includegraphics[width=0.95\linewidth]{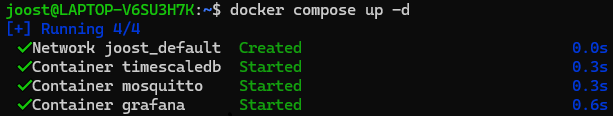}
    \caption{Docker compose startup.}
    \label{fig:docker-compose}
\end{figure}

\begin{figure}
    \centering
    \includegraphics[width=\linewidth]{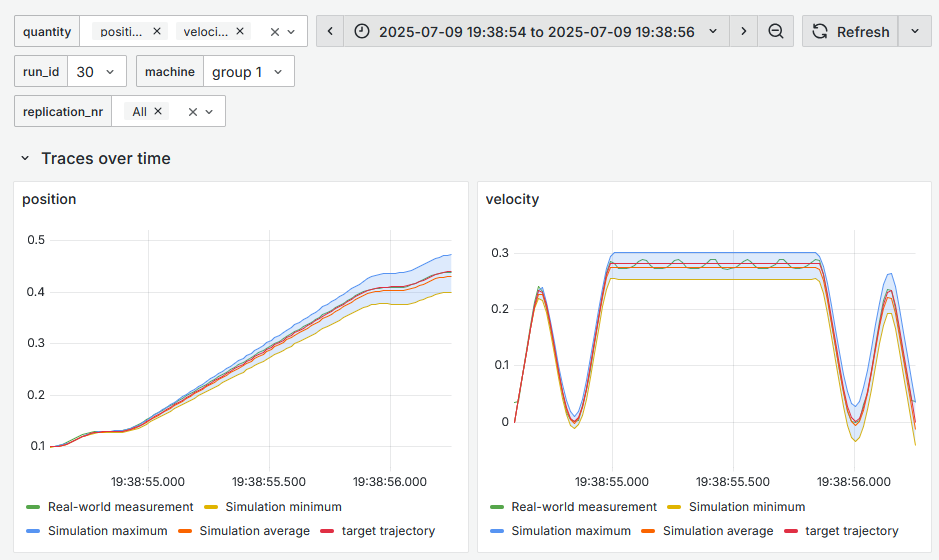}
    \caption{Grafana Dashboard showing trajectory traces.}
    \label{fig:dashboard}
\end{figure}

\Cref{fig:hmi,fig:docker-compose,fig:dashboard} show some results or output of this exemplar. \Cref{fig:hmi} shows the basic but functional Human Machine Interface the operator can use to control the crane. In it are controls for the crane's movement, homing and zeroing, the real-time status and configurations of the logger w.r.t database writeout. \Cref{fig:docker-compose} shows that with the single docker compose command the containers and the associated network are up and running. Lastly, \cref{fig:dashboard} shows some traces on the dashboard. Those traces have also been simulated by the simulator, as indicated by the upper (in blue) and lower (in yellow) confidence bounds derived from the simulation.

\subsection{Usage in Publications}

The first publication using the exemplar is \cite{Mertens-2023}, a short abstract on the concept of continuous validation. While the paper itself did not yet feature the gantry crane, the accompanying poster used the crane as illustrative example.

Later, in \cite{Mertens-2024}, we further explored the concept of continuous validation through the reuse of existing model validation techniques. Here the initial version of the crane exemplar was developed and used in the evaluation of the paper. For more detailled results w.r.t. the continuous validation digital twin we refer the reader to this publication.

The next use of the crane was in \cite{Mertens-2024b}, where we explored DarTwin, a notation for digital twin evolution. We identified multiple commonly used digital twin evolution. On the one hand the development process of the crane was used to derive those evolutions, on the other hand we envisioned multiple future evolutions of the crane system to demonstrate the DarTwin notation.

The latest use of the exemplar was in \cite{Mertens-2024a}, where we explored state-of-the-art time series classification techniques to find faults in systems in operation. In essence, we extended the continuous validation method of \cite{Mertens-2024} with classification techniques to better pinpoint system faults in the exemplar.

\subsection{Usage in Teaching}

At the University of Antwerp, we organized a Blended Intensive Programme (BIP) on the topic of Model Based Systems Engineering for Digital Twin Systems. Here the exemplar was used as project to work on. We expanded it to an entire harbor system, with a container identification system to intake containers in the harbor, and a ship digital twin with a buoyancy model to teach about correct container loading order. This BIP was organized during the first semester of academic year 2024-2025. In academic year 2025-2026 it will be revisited in the second semester.

At the Gran Sasso Science Institute (GSSI) the exemplar was used during the SpaceRaise doctoral school to teach about digital twinning in practice\footnote{\url{https://spaceraise.academy/\#courses} Week 2, May 19th 2025}. On this first day talks were given on the topic of twinning in Systems Engineering, and the exemplar was used to practically summarize the topics presented that day. Specifically for this doctoral school, the Capella models were developed.

\section{Discussion and Future Work}
\label{sec:discussion}

The gantry crane has proven it's merit for both research and education, but nonetheless it has some limitations or shortcomings which we discuss here.

\subsection{Cost}

Firstly, with the current BOM the crane is not low cost, at approximately 1250 Euro for one crane in parts. Almost half of this cost stems from the Trinamic Motor Drivers, for which a development kit is used. The logical next step is to forego the development kits and the measurement Arduino with its shield, and develop a custom driver board integrating these three microcontrollers. This would greatly bring the cost down.

\subsection{Size}

The current size of the crane was chosen based on the dimensions of the location in the lab. This size, however, is larger than the common size of checked luggage accepted by airlines, making it cumbersome to transport.

\subsection{Teaching Extensions}

During teaching we made extensions to the gantry crane: for the BIP more of the harbor was included, and for the doctoral school we focused on Capella models. We deemed those extensions a little too immature to include as features in this paper, in the future they must be further developed, especially the ship and harbor simulations for the BIP.

\subsection{Lack of Physical Simulation}

Currently, there is no physical crane simulator. To make the exemplar even more accessible, a completely simulated version of the physical crane could be included, as was done in the exemplars mentioned in the related work. This researchers or educators can test the project without committing to building the rather costly crane. This could also open up uses on the topic of virtual commissioning.

\subsection{Digital Twin Frameworks}

Over the past years a multitude of digital twin frameworks have been developed, many of them open-source \cite{Lee-2024,Gil-2024}. Examples are Eclipse Ditto, for Internet-of-Things devices and their digital twins, or Eclipse BaSyx for Asset Administration Shell driven digital twins. Our approach is much more ad-hoc, and more in line with the approach taken in \cite{Damjanovic-Behrendt-2019}. We make use of open-source technologies, but none of them are tailored specifically to digital twins. One could argue for or against this, in our opinion the ad-hoc approach allows more freedom and less overhead, but usage of the frameworks would surely make the exemplar more industry relevant.

\section{Conclusion}
\label{sec:conclusion}

We presented a lab-scale gantry crane exemplar designed to support digital twin research and education. It expands the amount of available digital twin exemplars and provides researchers and educators an off-the-shelf use case, reducing their development effort. To facilitate adoption and reproducibility, the exemplar is openly available on GitHub along with supporting documentation. The exemplar's utility was illustrated with examples from our personal academic publications and teaching use cases. While we deem the lab-scale gantry crane mature enough to be presented as exemplar, it still has some limitations, which we discussed and for which we provided pointers for future work.

\bibliographystyle{ieeetr}
\bibliography{bibliography}
\end{document}